\documentclass[apl,preprint]{revtex4-1}

\usepackage{graphicx}
\usepackage{dcolumn}
\usepackage{bm}
\usepackage{xcolor}
\usepackage{amsmath}

\begin{document}
\title{Experimental observation of plasmons in a graphene monolayer resting on a two-dimensional subwavelength silicon grating}
\author{Xiaolong~Zhu$^{1,2}$}
\author{Wei~Yan$^{1,2}$}
\author{Peter~Uhd~Jepsen$^1$}
\author{Ole~Hansen$^{3,4}$}
\author{N. Asger~Mortensen$^{1,2}$}
\author{Sanshui~Xiao$^{1,2}$}
\email{saxi@fotonik.dtu.dk}
\affiliation{$^1$ DTU Fotonik - Department of Photonics Engineering, \\Technical
University of Denmark, DK-2800 Kongens Lyngby, Denmark\\
$^2$Center for Nanostructured Graphene (CNG), Technical University of Denmark, DK-2800 Kongens Lyngby, Denmark \\
$^3$ DTU Nanotech - Department of Micro and Nanotechnology, Technical University of Denmark,
DK-2800 Kongens Lyngby, Denmark\\
$^4$ Center for Individual Nanoparticle Functionality (CINF), Technical University of Denmark,
DK-2800 Kongens Lyngby, Denmark}

\date{\today}

\begin{abstract}
We experimentally demonstrate graphene-plasmon polariton excitation in a continuous graphene monolayer resting on a two-dimensional subwavelength silicon grating. The subwavelength silicon grating is fabricated by a nanosphere lithography technique with a self-assembled nanosphere array as a template. Measured transmission spectra illustrate the excitation of graphene-plasmon polaritons, which is further supported by numerical simulations and theoretical prediction of plasmon-band diagrams. Our grating-assisted coupling to graphene-plasmon polaritons forms an important platform for graphene-based opto-electronics applications.
\end{abstract}
\pacs{42.79.Dj, 73.20.Mf, 78.66.Bz, 71.36.+c}
\maketitle

Graphene is a two-dimensional (2D) carbon-based material, whose unique electronic and optical properties have attracted a great deal of research interest.~\cite{Novo,Zhang,Grig,Bludov,Ferreira,Koppens} Novel opto-electronic applications such as waveguiding,~\cite{Xiao} ultrafast lasing,~\cite{Sunz} optical modulation,\cite{Lium} and photodetection~\cite{Fang} have been realized by effective doping and gating of graphene. Despite the fact that graphene is an atomically thin layer the optical absorption of a single layer can be as high as 2.3$\%$ (defined by the fine structure constant).~\cite{Mak2008P1} Nevertheless, for light-matter interactions this number is imposing challenges and restrictions for graphene-based opto-electronic devices.~\cite{Liuj} One promising way to enhance the optical absorption is to excite graphene-plasmon polaritons (GPPs) supported by graphene.
By periodical patterning of the graphene, excitations of GPPs in the structured graphene have been demonstrated both theoretically and experimentally, e.g., in one-dimensional micro-ribbons,~\cite{Niki,Jul} in 2D micro-disks,~\cite{Thon,Yanh} and in anti-dot~\cite{Nikit} arrays.
It is desirable to have a continuous sheet of graphene because it allows in-plane charge-carrier transport.
Recently, GPPs in a continuous graphene sheet have been experimentally detected using advanced near-field scattering microscopy with infrared excitation.~\cite{Feiz,Chen} Another method for excitation of GPPs in a continuous graphene sheet is to use silicon subwavelength gratings.~\cite{Zhan,Gaow} Besides, the silicon grating substrate could end up as a favorite choice in the monolithic integration of graphene-based optics and electronics in the future.~\cite{Baoq} In this Letter, we experimentally demonstrate the excitation of GPPs in a continuous graphene monolayer resting on a 2D subwavelength silicon grating (SWSG).
The SWSG is realized by a nanosphere lithography technique (a widely used technique for fabricating low-cost wafer-scale samples) with a self-assembled nanosphere array as a template.~\cite{Zhu,Zhux} A chemical-vapor-deposition-grown monolayer graphene on copper is then transferred onto the SWSG after chemically etching away the copper. Guided GPP resonances excited in the graphene monolayer on diffractive gratings are observed in transmission spectra measured at midinfrared frequencies. Numerical simulations and analytical resonance band properties are implemented to further support the experimental results. The experimental demonstrations of SWSG-assisted GPP excitation may open routes to build graphene-based plasmonic devices and circuits.

\begin{figure}[b]
\centering\includegraphics[width=10cm]{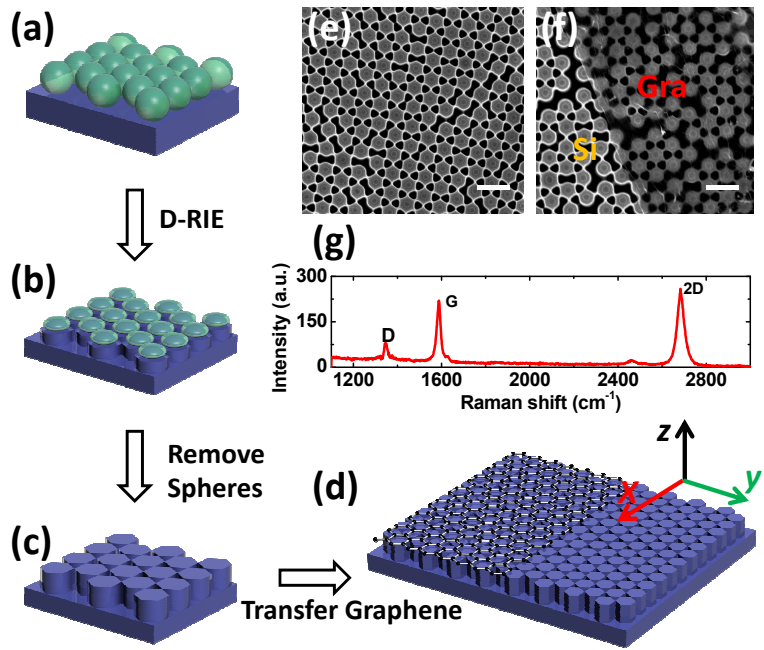}
\caption{ (Color online) (a)-(c) Fabrication processes of the subwavelength silicon grating structure. (d) Illustration of a silicon grating structure covered by a monolayer of graphene. Top-view SEM picture of the silicon grating (e) before and (f) after transfer of a monolayer graphene on top. The dark region to the right in (f) is graphene. The scale bars in (e) and (f) are 1\,$\mu$m. (g) Measured Raman spectrum of the transferred monolayer graphene on top of the subwavelength silicon grating, which shows characteristic peaks indicative of single-layer graphene. }\label{Fig1}
\end{figure}

The SWSG is fabricated in a single-step dry etching process by using a nanosphere array as a template. Figures 1(a) to 1(c) illustrate the detailed fabrication processes of the Si grating using nanosphere lithography. Firstly, a monolayer of hexagonal-close-packed polystyrene (PS) nanospheres with a diameter of 600 nm (size dispersion of 1\%) is formed on a pre-treated Si surface by a self-assembly method [see Fig.~\ref{Fig1}(a)]. Subsequently, the Si sample is subjected to deep reactive ion etching (D-RIE) for pattern transfer, where the self-assembled PS nanosphere monolayer serves as a shadow mask [see Fig.~\ref{Fig1}(b)]. Due to the isotropic nature of the chemical etching from the plasma gases and directional bombardment of accelerated ions from physical etching in RIE, the PS nanospheres are etched away in an anisotropic fashion, i.e. with different etching rates for the vertical and lateral directions. Consequently, the nanospheres become thinner while their transverse diameters decrease very slowly (especially in D-RIE process). A vertically aligned Si rod array is achieved as shown in Fig. ~\ref{Fig1}(c). Fig.~\ref{Fig1}(e) shows a top-view scanning electron microscope (SEM) image, where the Si rods are periodically arranged in a hexagonal lattice.

The next step is to transfer a monolayer graphene onto the fabricated SWSG substrate based on an improved graphene transfer process.~\cite{Lix} After releasing a para-Methoxy-N-methylamphetamine (PMMA) covered graphene by etching away copper and transferring the PMMA/Graphene stack onto the SWSG, an appropriate amount of liquid PMMA solution is dropped on the cured PMMA layer to fully dissolve the precoated PMMA. After mechanical relaxation the contact between graphene and the substrate is improved.  Afterwards, PMMA is dissolved in acetone. Fig.~\ref{Fig1}(d) is a schematic illustration of a graphene-covered Si grating and Fig.~\ref{Fig1}(f) is a SEM image of a representative sample. We find that the relatively large filling factor of the Si grating and the advanced transfer method leave the flattened graphene in a good condition, e.g., without pronounced buckling. To further explore the quality of our structure, a Raman spectrum of the transferred graphene on SWSG was measured, and spectrum shows peaks typical for single-layer graphene [see Fig.~\ref{Fig1}(g)]. It should be mentioned that the weak D-band intensity comes from wrinkled graphene regions where silicon rods are located.

\begin{figure}[t]
\centering\includegraphics[width=9 cm]{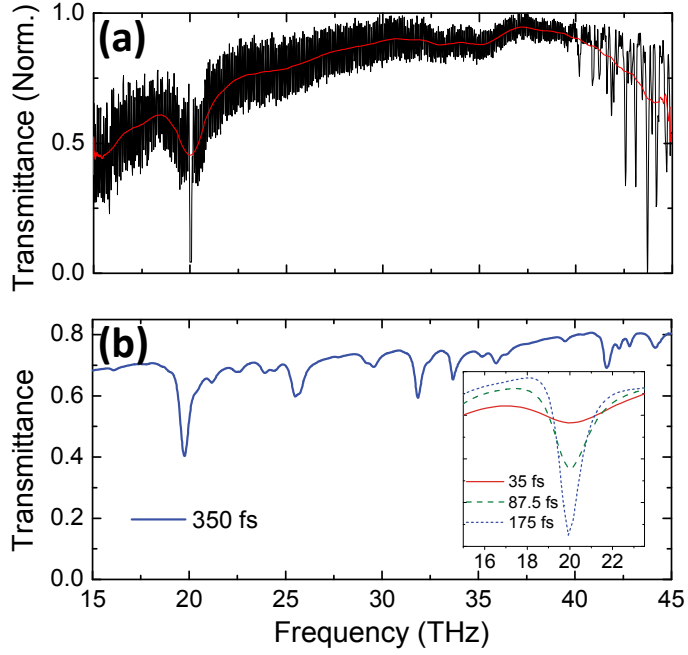}
\caption{ (Color online) (a) Transmittance spectrum of graphene-on-SWSG measured by Fourier transform infrared spectroscopy. The signals are normalized and average smoothing (red line) is used to illustrate the graphene plasmon resonance clearly. (b) Simulated transmittance spectrum. Here we choose $E_{\rm F}$ $\approx$ 0.35 eV and $\tau$ $\approx$ 350 fs. The inset shows how the resonance at around 20~THz varies at different relaxation times $\tau$.}\label{Fig2}
\end{figure}

We study the transmittance spectrum for the graphene resting on the SWSG using Fourier transform infrared spectroscopy (FTIR, Bruker VERTEX 70) at room temperature. Prior to measurements, the system was purged with nitrogen overnight to exclude the influence of water vapor in air. The unpolarized transmittance spectrum for the fabricated sample at midinfrared radiation from 15 THz to 45 THz is shown in Fig.~\ref{Fig2}(a), where the unpolarized transmittance for the bare SWSG sample is taken as a reference. The spectrum shows several prominent transmittance dips, e.g., at 20 THz (1 THz = 33.3 cm$^{-1}$) and 33 THz. In order to support the measured results, we performed full-wave numerical simulations based on a finite-integration technique (CST Microwave Studio).
The intraband conductivity of graphene is considered within the random-phase approximation.~\cite{Wuns,Hwan}
\begin{eqnarray}
\sigma (\omega ) &=& \frac{{2{e^2}}}{{\pi \hbar^2 }}\frac{i}{{\omega  + i\tau^{-1}}}k_BT\log \left[ {2\cosh \left( {{E_{\rm F}}/2{k_{\rm B}}T} \right)} \right].
\end{eqnarray}
Here, an intrinsic relaxation time $\tau=\mu E_{\rm F}/e\upsilon_{F}^{2}$ at room temperature is used, where $E_{\rm F}$ is the Fermi level, $\upsilon_{F}$ is the Fermi velocity, and $\mu$ is the dc mobility. The hexagonal rods array has a lattice period $\Lambda$ = 600 nm, an inner radius of hexagonal rods $r$ = 220 nm and a height of the rods $h$ = 490 nm (Surface profiler, Dektak), respectively. The simulated transmittance spectrum (average result of both polarizations) in Fig.~\ref{Fig2}(b) presents multiple dips, in a good agreement with the experimental one [Fig.~\ref{Fig2}(a)], especially for the pronounced dip at ~20 THz. The inset in the Fig.~\ref{Fig2}(b) shows how the resonance at around 20~THz mainly broadens when the relaxation time is reduced. One finds that the position of resonances associated with the GPP does not change, and that the resonance smears out as the relaxation time shortens, which is consistent with more general considerations. In this Letter, $E_{\rm F}$ $\approx$ 0.35 eV and $\tau$ $\approx$ 350 fs are used in the following simulations.

\begin{figure}[t]
\centering\includegraphics[width=10 cm]{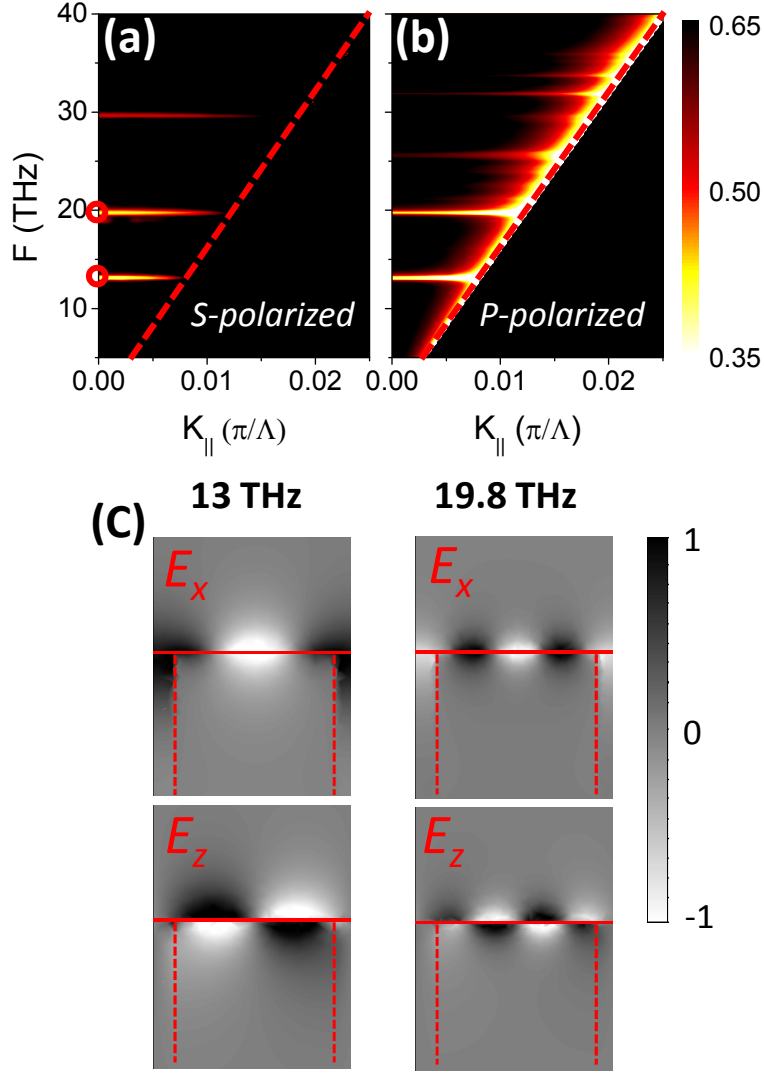}
\caption{ (Color online) Simulated transmittance spectra of the graphene-on-SWSG structure as functions of in-plane wave vector $\mathbf k_{||}$ and frequency for the (a) $s$ and (b) $p$ polarizations. The dashed lines in (a) and (b) indicate the light line.
(c) The electric field distributions of $E_{x}$ and $E_{z}$ for the fundamental- (13 THz) and second-order modes (19.8 THz) of the $s-$polarization. The dashed lines in (c) show the Si-air interfaces.}\label{Fig3}
\end{figure}

The dips observed in the transmittance spectra are associated to the excitation of the GPPs in the graphene layer.
Obviously, GPP modes cannot be excited directly in a homogeneous graphene sheet on a bare translationally invariant silicon substrate due to large wavevector mismatch. When the homogeneous graphene is periodically modulated by the SWSG, the GPP modes can be excited once the phase-matching condition is fulfilled, i.e.,
\begin{equation}
\mathbf k_{\rm GPP}=\mathbf k_{||}+\mathbf G,
\label{exct}
\end{equation}
where $\mathbf k_{\rm GPP}$ is the wave vector of the GPP mode and $\mathbf G$ is the reciprocal lattice vector of the grating.
To further illustrate the excitation of GPP modes, we simulated the transmittance spectra of the structure as functions of in-plane wave vector $\mathbf k_{||}$ and frequency for $s$-/$p$-polarization [electric/magnetic fields along the $x$ direction in Fig.~\ref{Fig1}(d)]. We find that the low-order resonance bands are excited efficiently. In particular, the second-order mode at 19.8 THz is well-excited for both polarizations, related to the deep transmittance dip at 20 THz in our measurement. Moreover, to excite the GPP mode more efficiently, it is preferred that the incident electric field has a component in the direction of $\mathbf k_{\rm GPP}$. Based on this consideration, the incident electric field for the $p$-polarization is more codirectional with $\mathbf k_{\rm GPP}$ than the case for the $s$-polarization, therefore making the $p$ wave more favorable for exciting the GPP mode. Additionally, the $p$ wave can have a component out of the grating plane for the oblique incident case, and such an out-of-plane component of the electric field also contributes to excite the GPP mode. This explains why we observe a rich spectrum with deeper dips for the $p$-polarized light [Fig.~\ref{Fig3}(b)].

The electric field distributions of $E_{x}$ and $E_{z}$ at the GPP
resonance are displayed in Fig.~\ref{Fig3}(c) for the low-order modes of the $s$-polarization.
From the electric field distributions of $E_{z}$ [bottom part of Fig.~\ref{Fig3}(c)], dipole- and quadrupole-like patterns are observed, which are fingerprints of the fundamental- and second-order GPP modes.
The apparent field enhancement near the graphene plane is a characteristic of GPPs. We find that the GPP modes are tightly
confined to the graphene plane, and the plasmon wavelength is about 40 times smaller than the dimension of the midinfrared wavelength. The deep subwavelength nature of GPPs can cause flat resonance bands above
the light line, which are almost unaffected when varying $\mathbf k_{||}$, as shown in Figs.~\ref{Fig3}(a) and~\ref{Fig3}(b).

\begin{figure}[htb]
\centering\includegraphics[width=12 cm]{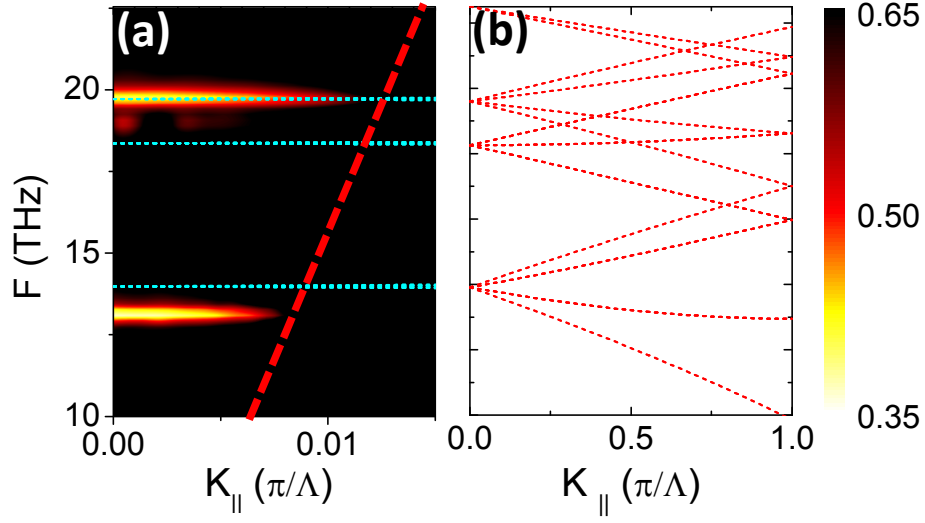}
\caption{ (Color online) (a) Simulated transmittance spectra of the graphene (part of the result shown in Fig. 3(a)). (b) Calculated plasmonic band diagram. The dotted lines in (a) show the superimposed plasmon bands from (b) with $\mathbf k_{\rm GPP}$ ranging from 0 to 0.015$\pi/\Lambda$, indicating almost completely flat GPP dispersion above the light line.}\label{Fig4}
\end{figure}

Since the periodicity of the grating is much smaller than the wavelength of the incident wave, we can approximately model the grating as an effective medium with the permittivity $\epsilon_{\rm eff}=f_{\rm silicon}\epsilon_{\rm silicon}+(1-f_{\rm silicon})\epsilon_{\rm air}$, with the filling factor $f_{\rm silicon}$ = 0.49 for our specific structure. Then, the wavenumber of the GPP mode is approximated as
\begin{equation}
k_{\rm GPP}\approx(\epsilon_{\rm eff}+1)\frac{i\omega\epsilon_0}{\sigma}.
\label{ksp}
\end{equation}
The band structure of the GPP mode, i.e., the $\mathbf k_{||}$-$\omega$ relation, can be approximately extracted from Eqs. (\ref{exct}) and (\ref{ksp}). For the hexagonal lattice grating composed of silicon, the lowest few bands are plotted in Fig.~\ref{Fig4}(b) with $\mathbf k_{||}$ varying along the ${\Gamma \rm M}$ direction in the Brillouin zone. It is observed that the GPP bands qualitatively agree well with the positions of the absorption peaks, i.e, transmission dips in Fig.~\ref{Fig4}(a) for the $s$-polarized case. Note that Fig.~\ref{Fig4}(a) is part of the results shown in Fig.~\ref{Fig3}(a).
This further indicates that the enhanced absorption in the graphene structure results from the excitation of the GPP mode. In particular, the excited bands around 13.8 THz and 19.8 THz in Fig.~\ref{Fig4}(b) correspond to $\mathbf G$ in the ${\Gamma \rm M}$ direction, while the dark bands around 18.2 THz are for $\mathbf G$ in the ${\Gamma \rm K}$ direction. The existence of the dark bands is attributed to the symmetry mismatch between the incident plane wave and the Bloch wave.~\cite{Fans}

In conclusion, we have experimentally demonstrated SWSG-assisted GPP excitation in a continuous graphene monolayer. The SWSG is fabricated based on low-cost nanosphere lithography. An improved method for transferring CVD-graphene on copper is used to obtain a high quality of the transferred graphene monolayer. With the aid of the SWSG, the excitation of GPP resonances has been observed in the measured FTIR spectrum, confirmed by numerical simulations and theoretical predictions of the plasmon-band relations. Our approach forms a basis for a paradigm in optical and opto-electronic detectors and light-harvesting devices. By applying proper doping or gating, our structure can also inspire the development of active-tunable filters and modulators.

$\\$
This work is partly supported by the Catalysis for Sustainable Energy Initiative Center, funded by the Danish Ministry of Science, Technology and Innovation.
The Center for Nanostructured Graphene (CNG) is sponsored by the Danish National Research Foundation, Project DNRF58. The Center for Individual Nanoparticle Functionality CINF is funded by the Danish National Research Foundation (DNRF54). We thank P.~B{\o}ggild (DTU Nanotech) for providing us access to Raman spectroscopy.


%

\newpage
\section{Figure captions}
\textbf{Figure 1}: (Color online) (a)-(c) Fabrication processes of the subwavelength silicon grating structure. (d) Illustration of a silicon grating structure covered by a monolayer of graphene. Top-view SEM picture of the silicon grating (e) before and (f) after transfer of a monolayer graphene on top. The dark region to the right in (f) is graphene. The scale bars in (e) and (f) are 1\,$\mu$m. (g) Measured Raman spectrum of the transferred monolayer graphene on top of the subwavelength silicon grating, which shows characteristic peaks indicative of single-layer graphene.

\textbf{Figure 2}: (Color online) (a) Transmittance spectrum of graphene-on-SWSG measured by Fourier transform infrared spectroscopy. The signal is normalized and average smoothing (the red line) is used to illustrate the graphene plasmon resonance clearly. (b) Simulated transmittance spectrum. Here we choose $E_{\rm F}$ $\approx$ 0.35 eV and $\tau$ $\approx$ 350 fs. The inset shows how the resonance at around 20~THz varies at different relaxation times $\tau$.

\textbf{Figure 3}: (Color online) Simulated transmittance spectra of the graphene-on-SWSG structure as functions of in-plane wave vector $\mathbf k_{||}$ and frequency for the (a) $s$ and (b) $p$ polarizations. The dashed lines in (a) and (b) indicate the light line.
(c) The electric field distributions of $E_{x}$ and $E_{z}$ for the fundamental- (13 THz) and second-order modes (19.8 THz) of the $s-$polarization. The dashed lines in (c) show the Si-air interfaces.

\textbf{Figure 4}: (Color online) (a) Simulated transmittance spectra of the graphene (part of the result shown in Fig. 3(a)). (b) Calculated plasmonic band diagram. The dotted lines in (a) show the superimposed plasmon bands from (b) with $\mathbf k_{\rm GPP}$ ranging from 0 to 0.015$\pi/\Lambda$, indicating almost completely flat GPP dispersion above the light line.

\newpage
\clearpage

\newpage
\clearpage

\newpage
\clearpage

\newpage
\clearpage
\end{document}